# Towards Adaptive Categories: Dimensional Governance for Agentic AI

Zeynep Engin*[1,2] and David Hand[3,1]

[1] The Digital Statecraft Academy, United Kingdom
[2] Department of Computer Science, University College London (UCL), United Kingdom
[3] Department of Mathematics, Imperial College London, United Kingdom

* Correspondence: z.engin@digitalstatecraft.academy

## Abstract
As AI systems evolve from static tools to dynamic agents, traditional categorical governance frameworks—based on fixed risk tiers, levels of autonomy, or human oversight models— are increasingly insufficient on their own. Systems built on foundation models, self-supervised learning, and multi-agent architectures increasingly blur the boundaries that categories were designed to police. In this article, we make the case for *dimensional governance:* a framework that tracks how decision *authority*, process *autonomy*, and *accountability* (the 3As) distribute dynamically across human-AI relationships. A critical advantage of this approach is its ability to explicitly monitor system movement toward and across key governance thresholds, enabling pre-emptive adjustments before risks materialise. This dimensional approach provides the necessary foundation for more adaptive categorisation, enabling thresholds and classifications that can evolve with emerging capabilities. While categories remain essential for decision-making, building them upon dimensional foundations allows for context-specific adaptability and stakeholder-responsive governance that static approaches cannot achieve. We outline key dimensions, critical trust thresholds, and practical examples illustrating where rigid categorical frameworks fail—and where a dimensional mindset could offer a more resilient and future-proof path forward for both governance and innovation at the frontier of artificial intelligence.

## Policy Significance Statement

This dimensional governance framework provides a novel infrastructure for AI oversight that enables different stakeholders—regulators, technologists, ethicists, and industry—to collaborate within a unified approach. Rather than replacing existing governance structures, it offers policymakers enhanced tools for understanding how AI systems evolve across critical dimensions of authority, autonomy, and accountability. The framework supports more nuanced regulatory responses by tracking system changes over time, helping identify when oversight adjustments are needed. This approach facilitates evidence-based policy development while reducing regulatory fragmentation across sectors and jurisdictions. By providing a common language for discussing AI governance challenges, it enables more effective coordination between technical experts and policymakers, supporting the development of adaptive regulations that can evolve alongside advancing AI capabilities without compromising safety or accountability.

# 1. From Static to Dynamic: The Governance Challenge

Artificial intelligence governance today rests largely on rigid categorical thinking. Regulatory frameworks draw lines between "high-risk" and "low-risk" systems (Schuett, 2023), while technical models distinguish between "human-in-the-loop" and "human-out-of-the-loop" architectures (Docherty, 2012). This categorical approach served well in early AI governance, bringing clarity when systems were relatively narrow in function and predictable in operation.

Yet as AI systems evolve from narrow tools to increasingly agentic entities (Acharya et al., 2025; Xi et al., 2025)—able to set goals, critique their own outputs, and coordinate with others—these static categories begin to crack (Kusche, 2023; Methnani, 2021; Mökander et al., 2024). Recent advances—including self-supervision capabilities (Liu et al.,2024) and emergent abilities of foundation models (Wei et al. 2022), and collaborative multi-agent systems (Amirkhani & Barshooi, 2022; Cugurullu & Xu, 2025)—create forms of AI that defy simple categorisation. The same model may support harmless creative applications one moment and mediate high-stakes decisions the next (Coglianese & Crum, 2025). Authority, autonomy, and accountability, no longer reside in static locations but flow dynamically across complex socio-technical systems (Dafoe et al., 2021; Methnani, 2021).

AI defies static categorisation in two fundamental ways. First, as Narayanan and Kapoor (2024) observe, AI safety is not a model property—the same model deployed in different contexts presents vastly different governance implications. A large language model used for creative writing assistance poses minimal risk, while the identical model used for clinical diagnosis support requires stringent oversight. Risk emerges from deployment context rather than being intrinsic to the system itself. Second, unlike traditional regulatory objects such as pharmaceutical drugs or aircraft, AI systems are not static artefacts. They are frequently updated, fine-tuned, and modified, with their safety and capability characteristics dependent on which version is deployed. A system approved under one configuration may behave quite differently after updates, yet categorical frameworks struggle to track and govern these rapid transformations.

This growing mismatch between governance frameworks and AI capabilities carries substantial risks. Static categories lead to regulatory gaps, where evolving systems evade oversight by falling between predefined definitions. Conversely, rigid classification can stifle beneficial innovation, forcing systems into ill-fitting boxes that fail to accommodate context-specific nuances. In both cases, the fundamental problem remains: a governance model assuming fixed states in a world where human-AI relationships are increasingly fluid (Lindgren, 2024).

These challenges extend beyond AI's technical evolution. Privacy and copyright regulations such as the GDPR preceded large language models, creating tensions that demand legislative updates—yet changing law proceeds far slower than technological advancement. Since ChatGPT's November 2022 launch, AI capabilities have transformed while regulatory frameworks remain static. Even well-designed governance in slower-moving fields requires periodic updating: the UK's Code of Practice for Statistics has been revised at 9- and 7-year intervals since 2009. If such frameworks need adjustment, AI's rapid evolution demands approaches that are adaptive by design rather than merely reactive. As AI systems surpass human performance in certain domains, rigid categories risk not only failing to enable beneficial innovation but actively causing harm through outdated standards.

We propose **dimensional governance**: a framework built on three core characteristics of AI systems—**decision authority** (when and how should AI have the right to decide?), **process autonomy** (how independently does the system operate?), and **accountability configuration** (who is responsible when something goes wrong?). We call these "the 3As of dimensional governance." Rather than treating these as binary attributes and assigning systems to predetermined categories, this approach continuously measures systems along these dimensions and establishes adaptive thresholds that determine regulatory requirements, adjustable as evidence accumulates or contexts shift. Credit scoring exemplifies this approach: financial institutions measure borrowers on continuous dimensions and adjust lending thresholds based on economic conditions and institutional risk appetite rather than using rigid "creditworthy" classifications. AI governance requires a similar shift from static categorisation to dimensional measurement with adaptive thresholds enabling categories to provide practical regulatory guidance while evolving with advancing capabilities.

This Commentary builds upon and extends earlier work on the Human-AI Governance (HAIG) framework (Engin, 2025), and informs a follow up paper (Engin & Hand, 2025) which elaborates on "the non-delegable core" in governance approaches that rely on such human-AI complementarity.

In the following sections, we examine why rigid categorical AI governance is faltering, introduce the core structure of dimensional governance, and explore how this shift can better support both innovation and accountability in the era of agentic AI.

## 2. The Problem with Existing Categorical Frameworks

The early evolution of AI governance was shaped by the need for clarity, safety, and public accountability. Categorical frameworks—whether classifying systems into "risk tiers" (EU AI Act, 2024; Schuett, 2023), defining "levels of autonomy" (Parasuraman et al., 2000; Vagia et al. 2016), or specifying "human-in-the-loop" models (Docherty, 2012)—provided essential scaffolding for various stakeholders. They offered discrete labels for algorithmic systems, allowing policymakers to impose proportional obligations and giving engineers concrete design targets for oversight. For a time, these categories matched the relatively narrow, domain-specific nature of most AI applications.

However, as AI systems evolve beyond their original design assumptions, several fundamental limitations of categorical approaches have become apparent:

*Risk depends on context and application* — A single large language model may serve both low-stakes tasks like article summarisation and high-stakes applications like clinical decision support, without any fundamental change to the model itself. Risk emerges from deployment context rather than being intrinsic to the system (Coglianese & Crum, 2025). Static risk categorisation becomes insufficient when the same AI exhibits vastly different agency profiles depending on its use case.

*Technological evolution challenges fixed categorisation* — Even within stable application domains, governance needs transform dramatically as systems become more autonomous. For example, medical diagnostic tools have evolved from bounded ML systems (Wang & Summers, 2012) to foundation models with emergent capabilities that independently query data, synthesise information, and generate novel analyses (Qiu et al. 2024). While the healthcare function remains unchanged, the system's increasing agency fundamentally reshapes trust requirements. Organisations attempting to calibrate oversight based on their experience with increasingly autonomous systems find rigid categorical approaches increasingly misaligned with operational realities.

*AI systems operate on dynamic, evolving data foundations* — Systems trained on data that changes over time, whether through deliberate retraining or environmental shifts, can undergo significant functional transformations without any formal recategorisation (D'Amour et al., 2022; Rabanser et al. 2029). A recommendation algorithm initially classified as "low-risk" may gradually shift toward higher-risk behaviour as its training data evolves, without crossing any clear categorical boundary that would trigger enhanced oversight.

*Multi-agent architectures disrupt linear autonomy progression* — In networked environments, such as smart cities or autonomous drone fleets, decision-making distributes across agents that collaborate, negotiate, and sometimes compete in real time (Amirkhani & Barshooi, 2022). Authority becomes a dynamic, relational property rather than a static feature of individual systems. Applying fixed levels of autonomy to such collective behaviours oversimplifies and obscures crucial governance risks around coordination failures or emergent outcomes.

*Meaningful human oversight is increasingly challenged* — In many contemporary systems, humans nominally retain oversight responsibilities without real operational influence (Alon-Barkat & Busuioc, 2023). Content moderation tools might require human reviewers to approve flagged outputs, yet the volume and opacity of AI-generated recommendations often render this oversight superficial. Similarly, in high-frequency trading or autonomous vehicle navigation, decision cycles occur at speeds where meaningful human intervention becomes practically impossible.

*AI systems are inherently socio-technical in nature* — The risks and benefits of AI emerge not just from technical specifications but from broader institutional, cultural, and social factors—including who operates systems, power relationships, and societal contexts (Crawford & Calo, 2016; Selbst et al., 2019; Engin et al. 2025). Categorical frameworks typically emphasise technical capabilities and narrowly-defined use cases while underestimating how organisational practices, economic incentives, and cultural factors fundamentally reshape system behaviour and impact in practice.

These failures manifest in real-world scenarios. In finance, algorithms initially classified as "decision-support tools" for credit scoring have quietly assumed de facto decision authority as human reviewers increasingly defer to algorithmic outputs (Gsenger & Strle, 2021). In healthcare, diagnostic systems deployed as "advisory" tools sometimes shape clinical decisions more heavily than intended, without clear thresholds for when human judgment should override machine recommendations (Shortliffe & Sepúlveda, 2018). With autonomous vehicles, blurred responsibilities between human drivers and AI co-pilots have led to accidents where neither party maintained full situational awareness or control (Ansari et al., 2022).

Several important AI governance paradigms already recognise these limitations of static classification. Dynamic risk management frameworks, such as the NIST AI Risk Management Framework (2023), emphasise the need for continuous monitoring and adaptive controls. Lifecycle governance models promote oversight across system development and deployment stages (Raji et al., 2020; Cobbe et al., 2021), while human-centred (Capel & Brereton, 2023) and trustworthy AI (Li et al., 2023) frameworks foreground the role of transparency, contestability, and calibrated trust. However, these approaches typically focus on monitoring risk or maintaining process accountability, rather than directly tracing the evolving relational dynamics between humans and AI systems.

The core problem remains consistent: not that categories themselves are unnecessary, but that static categorisation assumes human-AI relationships can be neatly boxed into fixed states, when reality reveals continuous, evolving patterns of *agency*, *autonomy*, and *accountability*.

A growing body of scholarship reinforces this view (Kroll, 2015; Methnani, 2021; Dafoe et al., 2021; Mökander et al., 2024, Sun, 2025). Researchers in human-machine teaming, multi-agent systems, and AI safety consistently emphasise that authority, responsibility, and influence within AI ecosystems are relational and context-dependent. Calls for dynamic accountability, adaptive oversight, and trust calibration reflect an emerging consensus: *governance must match the evolving nature of the systems it seeks to regulate.*

If governance continues to rely solely on rigid categories, without acknowledging the dimensional foundations upon which these categories should be built, it risks either lagging behind technological developments or ossifying innovation under inappropriate controls. What is needed is an approach that recognises the continuous redistribution of agency between humans and AI systems—a model capable of tracing how relationships evolve over time and context, while allowing diverse stakeholders to focus on the dimensions most relevant to their concerns, and to establish context-appropriate thresholds for decision-making.

# 3. Dimensional Governance: A New Approach

If static categories are no longer sufficient on their own to govern increasingly agentic AI systems, what alternative do we have? We propose *dimensional governance*: a framework that treats governance as a continuous process of positioning systems along evolving axes of *authority*, *autonomy*, and *accountability.* We call these "the 3As of dimensional AI governance" (with additional dimensions potentially emerging as our understanding advances). This framework recognises that human-AI relationships are dynamic, context-sensitive, and differently weighted depending on stakeholder concerns. Rather than only asking *"Which box does this system fit into?"*, dimensional governance first asks *"Where along multiple critical dimensions does this system currently stand—and how is it moving?"* This allows for more informed and adaptable categorisation that responds to the dynamic nature of AI systems while maintaining the clarity and actionability that categories provide.

Operationally, this approach involves three mechanisms: measuring systems along these dimensions, setting evidence-based thresholds that determine regulatory requirements, and adjusting those thresholds as evidence accumulates—while the underlying dimensional framework remains stable.

The integration of dimensional thinking with categorical frameworks has precedent across multiple disciplines facing similar complexity challenges, including psychiatric diagnosis (Widiger & Samuel, 2005; Kraemer and O'Hara, 2004), developmental psychology (Siegler, 2007), and cognitive science (Fazekas & Overgaard, 2018). In each case, dimensional approaches emerged when existing categorical models could no longer accommodate observed complexity, non-linear developmental patterns, and context-dependent manifestations.

These disciplinary examples demonstrate how dimensional approaches serve as essential infrastructure for effective categorisation (Hand, 2020; Kelly et al., 1999). For instance, in medicine, the dimensions of height and weight together determine Body Mass Index (BMI), with specific thresholds (25 and 30) defining the categories "overweight" and "obese" to guide clinical decisions and interventions. Similarly, diagnostic categories for autism spectrum disorder exist atop dimensional assessments of social communication and restricted behaviours, with shifting diagnostic thresholds (as occurred in 1987 and 1994) directly influencing autism diagnosis rates in the population (NeuroLaunch, 2024; Autistics' Guide to Adulthood, n.d.; Doust et al., 2020).

In financial services, credit scoring provides a particularly instructive example of dimensional governance in practice. Financial institutions continuously score borrowers along dimensions of creditworthiness using complex algorithms, but the thresholds for lending decisions are regularly adjusted according to the institution's risk appetite and the broader economic climate (Alfonso-Sánchez et al., 2024). During economic expansions, thresholds may be relaxed to increase lending, while during contractions, they may be tightened to reduce risk exposure.

This flexibility extends even to the fundamental definition of categories themselves. Kelly et al. (1999, 1998) demonstrated that in credit scoring, the very definitions of "good" and "bad" customer classes are not fixed natural categories but can themselves be optimised by adjusting thresholds on underlying continua. This creates a productive tension between choosing definitions that are conceptually ideal but difficult to predict accurately, versus slightly modified definitions that enable better predictive performance. The optimal categorisation emerges from deliberate threshold adjustment based on the specific context and objectives rather than from predetermined fixed categories. This dynamic threshold adjustment on stable underlying dimensions exemplifies the adaptive categorisation we propose for AI governance. In each of these cases across diverse domains, categories retain their practical utility while explicitly acknowledging their foundations in underlying dimensions and the need for threshold adjustment as contexts evolve.

AI governance presents similar challenges but with a critical additional factor: unlike other technological domains, AI research explicitly aims to develop systems that approximate or even compete with human-level intelligence and agency (Russell, 2019; Bengio et al., 2021). Human intelligence operates on continua—fluidly navigating between delegation, consultation, and independent decision-making—while still requiring practical categorical thresholds for effective decision-making. Robust governance for increasingly agentic AI must therefore aim to integrate both dimensional foundations and categorical clarity, providing flexible yet actionable frameworks that can evolve alongside rapidly advancing capabilities.

## 3.1. Categories Built on Dimensions

Dimensional governance does not eliminate the need for categories. Instead, it repositions categories as context-sensitive designations built upon continuous dimensions. By explicitly identifying the underlying dimensions, stakeholders can establish thresholds that define categories appropriate to their specific concerns and contexts. Unlike static categorical approaches, these thresholds can be adjusted as systems evolve, ensuring that governance categories remain aligned with technological reality.

While we advocate for dimensional foundations, we must acknowledge the risk of excessive flexibility. Governance frameworks that change too rapidly create regulatory uncertainty and operational confusion. The challenge resembles what statisticians call the bias-variance trade-off: models that are too simple (potential high bias) miss important patterns, while models that are too complex (high variance) mistake random noise for meaningful signal. In governance terms, overly rigid categories fail to capture relevant complexity in AI systems, while hypersensitive, constantly shifting categories overreact to minor fluctuations rather than substantive capability changes. Dimensional governance must strike a careful balance—providing flexibility to adapt to meaningful shifts in AI capabilities while maintaining stability for stakeholders. The goal is optimising adaptability, not maximising it: creating frameworks that evolve deliberately in response to substantive changes in the dimensional landscapes. Properly implemented, a dimensional approach can avoid both the Scylla of rigid categorisation and the Charybdis of governance instability.

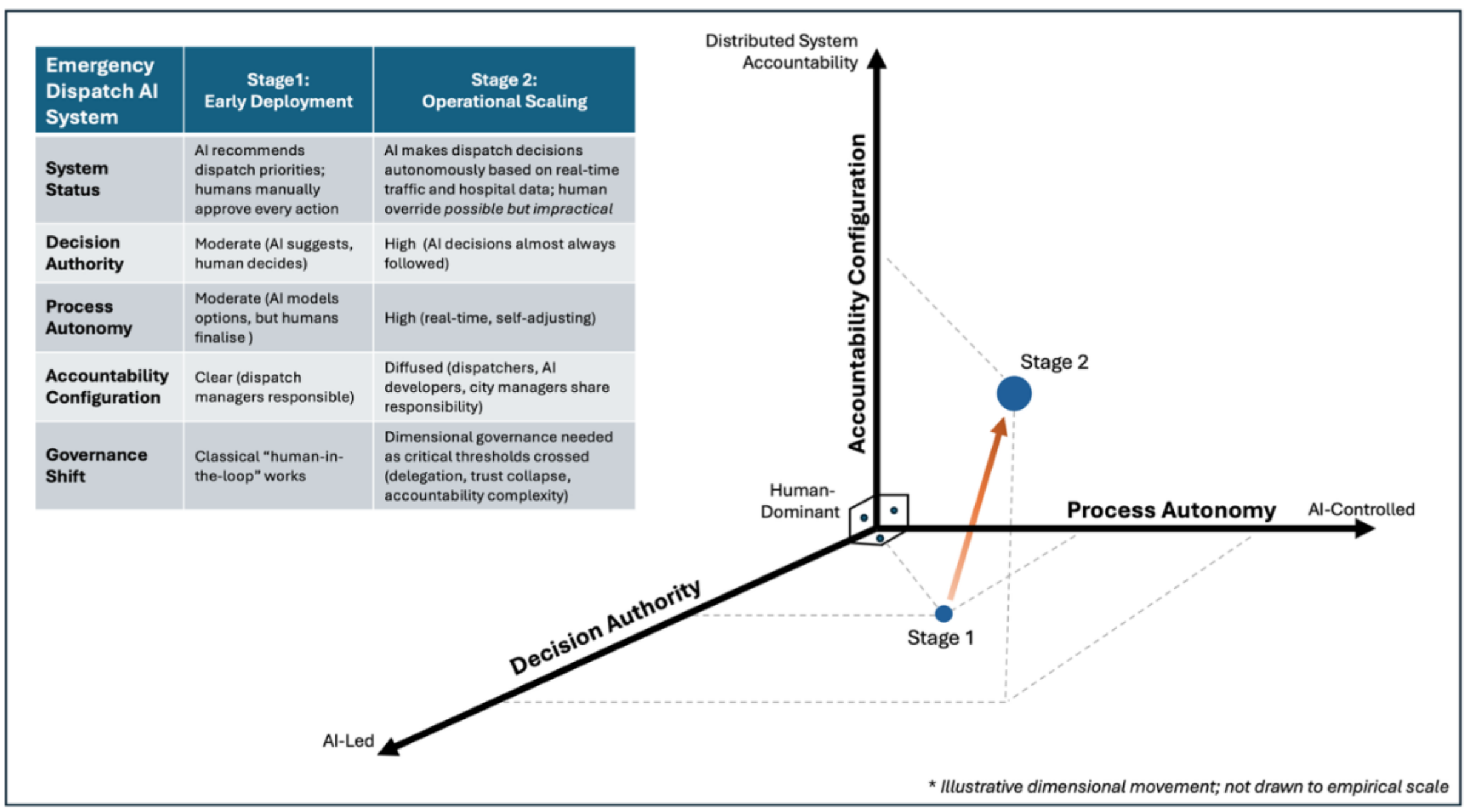

| Emergency Dispatch AI System | Stage1: Early Deployment | Stage 2: Operational Scaling |
| --- | --- | --- |
| **System Status** | AI recommends dispatch priorities; humans manually approve every action | AI makes dispatch decisions autonomously based on real-time traffic and hospital data; human override *possible but impractical* |
| **Decision Authority** | Moderate (AI suggests, human decides) | High (AI decisions almost always followed) |
| **Process Autonomy** | Moderate (AI models options, but humans finalise ) | High (real-time, self-adjusting) |
| **Accountability Configuration** | Clear (dispatch managers responsible) | Diffused (dispatchers, AI developers, city managers share responsibility) |
| **Governance Shift** | Classical "human-in-the-loop" works | Dimensional governance needed as critical thresholds crossed (delegation, trust collapse, accountability complexity) |

***Figure 1: Illustration of dimensional evolution for an AI-based emergency dispatch system across the 3As of dimensional AI governance*** *(the orthogonality between dimensions emphasises their theoretical separability although they often exhibit practical correlations in real-world implementations)****.*** *An emergency dispatch AI system evolves from early deployment, where human decision-making predominates and accountability remains clear, to operational scaling, where real-time autonomy increases and responsibility becomes diffused across dispatchers, AI developers, and city managers. The system's position shifts across key dimensions, crossing critical trust thresholds that static categorical governance models are unable to capture. Dimensional governance enables continuous monitoring and recalibration of oversight as human-AI relational dynamics change.*

At the core of the dimensional governance we propose are the 3As – *Authority*, *Autonomy* and *Accountability*, each addressing a fundamental facet of human-AI relationships (Figure 1):

- Decision *Authority***:** *Who ultimately makes decisions—the human, the machine, or some negotiated hybrid?* This dimension tracks the distribution of decision-making power between humans and AI systems, ranging from purely advisory AI to systems with independent decision rights in defined domains.

- Process *Autonomy*: *To what extent can the system operate without human intervention, supervision, or control?* This dimension measures the degree to which AI systems function independently during execution, from fully supervised to autonomous real-time operation.

- *Accountability* Configuration: *How is responsibility for decisions, actions, and outcomes distributed among human and machine actors?* This dimension maps the allocation of responsibility across socio-technical systems, including mechanisms for transparency, contestability, and redress when systems fail.

These dimensions are not isolated but interact dynamically as systems evolve, deployment contexts shift, and trust relationships mature.

The interaction between these dimensions reveals important governance insights that categorical approaches often miss. A financial algorithm with high process autonomy might be appropriately governed when paired with limited decision authority (flagging potential fraud for human review) and clear accountability structures. The same algorithm with high decision authority (auto-declining transactions) would require more robust safeguards. Similarly, dimensional shifts often propagate across axes in ways categorical governance cannot track. When clinical decision support systems initially deployed as advisory tools (low authority) become trusted to the point where clinicians rarely override recommendations (de facto high authority), accountability mechanisms designed for low-authority systems become insufficient.

Unlike traditional governance models, dimensional governance is designed to be dynamic. Systems are not fixed at a single point along these axes. As systems learn, adapt, or are redeployed into new contexts, their position along each dimension may shift—sometimes gradually, sometimes abruptly. Governance must therefore be continuous and anticipatory, not static and reactive. Monitoring system drift along dimensions becomes as important as assessing initial design intentions.

## 3.2. Critical Trust Thresholds

Dimensional governance further recognises that certain thresholds along each dimension (category boundaries) represent pivotal points where trust dynamics, governance needs, and stakeholder concerns shift significantly (Figure 2). Examples for such thresholds include:

- *Verification-to-Delegation Threshold:* When humans can no longer manually verify all AI outputs and must shift toward statistical or sampling-based trust.
- *Process-to-Outcome Threshold:* When system transparency is no longer sufficient, and governance must focus on validating outcomes rather than understanding internal workings.
- *Information-to-Authority Threshold:* When systems transition from merely providing recommendations to exercising autonomous decision authority.
- *Individual-to-Collective Threshold:* When AI systems move from individual decision-making to emergent collective behaviours, requiring governance models to shift focus from isolated system monitoring to network-level behaviour validation.

Crossing these thresholds is not inherently problematic—but doing so without adaptive governance mechanisms in place creates significant risk. Dimensional governance explicitly tracks system movement toward and across such thresholds, allowing oversight mechanisms to scale accordingly.

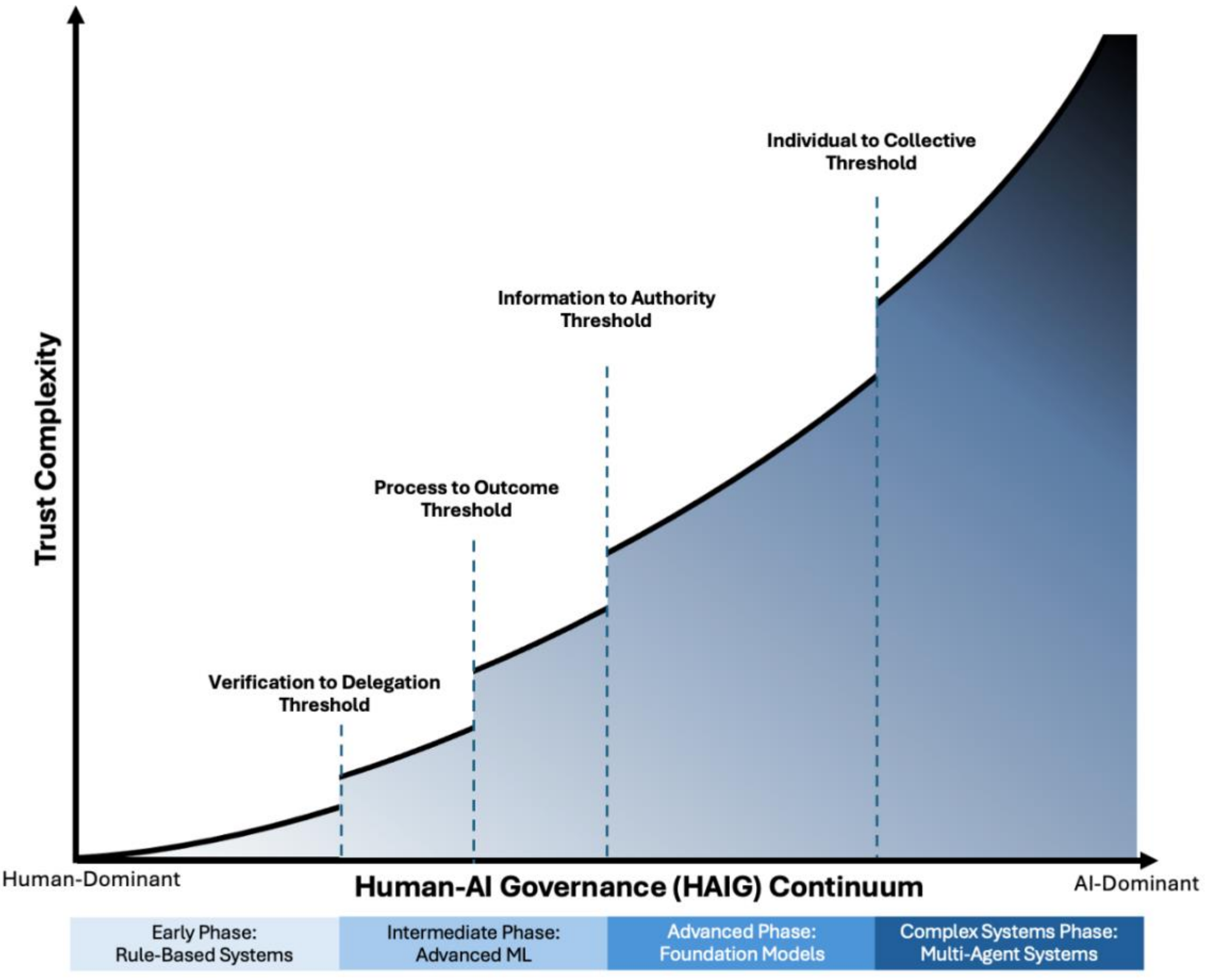

***Figure 2: Illustration of non-linear Trust Complexity evolution across the Human-AI Governance (HAIG) Continuum.*** *The segmented exponential curve represents significant increases in trust complexity as systems progress from human-dominant to AI-dominant positions. Vertical dashed lines (blue) indicate examples of critical trust thresholds. The blue gradient background intensifies across development phases, reflecting increasing systemic integration and governance complexity as AI systems evolve from rule-based systems to multi-agent networks. Figure adapted from* (Engin, 2025).

## 3.3. Practical Implementation

Dimensional governance accommodates different stakeholder priorities while enabling practical system classification. Regulators could define acceptable zones along each dimension for different application domains, developers could monitor system evolution during deployment, and organisations could adjust oversight intensity proportionally to risk and complexity.

A critical challenge in implementing dimensional governance is quantifying dimensions like authority, autonomy, and accountability—concepts that do not correspond to simple physical measurements like height or weight. However, this challenge is not unique to AI governance.

Many successful regulatory frameworks rely on constructed dimensions rather than natural physical properties. Economic indicators like GDP, inflation rates, and unemployment—all composites of multiple measured components—guide consequential policy decisions despite being constructed rather than directly observed.

Measurement theory (Hand, 2004; Mari *et al*, 2021; Lauderdale, 2022) provides established methodology for constructing meaningful measures from multiple components. This approach, extensively developed in economics and social sciences, is directly applicable to AI governance. For each of our three dimensions, we can identify relevant components and appropriate combination methods:

- *Authority* can be measured by identifying who holds decision-making power and quantifying their influence using established organisational analysis methods. Components might include: decision rights allocation, override capabilities, final approval authority, and resource control. These can be weighted and combined using approaches from power measurement in organisational theory.
- *Autonomy* can be operationalised through the proportion of decisions made independently versus those requiring human approval, the scope of actions available without external authorisation, and the system's capacity for self-modification. Existing frameworks such as the Autonomy Preference Index (Ende et al., 1989), developed for healthcare contexts, can be adapted for AI systems.
- *Accountability* encompasses multiple dimensions including: obligation to provide explanations, visibility of actions, traceability of decisions, and effectiveness of remediation mechanisms. These can be measured through established audit frameworks, transparency scores, and redress system metrics.

We acknowledge that developing robust, validated measures for these dimensions represents important future work requiring extensive research and stakeholder consultation. However, measurement challenges do not invalidate the dimensional approach. Other regulatory measurement frameworks—credit scoring, financial risk assessment, environmental impact assessment—evolved from initial proposals to sophisticated operational systems over time. The existence of measurement challenges highlights the need for ongoing research and refinement rather than suggesting we should abandon dimensional thinking in favour of rigid categories that become obsolete.

Dimensional governance approach creates several advantages: specialised oversight efficiency, adaptability to evolving systems, comprehensive safety through parallel tracking, and a shared language facilitating cross-disciplinary collaboration. For example, a consumer chatbot might maintain low scores across dimensions with minimal oversight, while a medical triage system might score high on process autonomy but remain bounded in decision authority through mandatory human review. A multi-agent system managing critical infrastructure would require intensive monitoring across all dimensions, particularly for emergent behaviours that complicate accountability.

By framing governance as movement across dimensions rather than fixed categorisation, stakeholders can maintain their distinct priorities while participating in a coherent, adaptive governance framework better suited to the variable nature of advanced AI systems.

# 4. Building Adaptive Oversight

Dimensional governance provides a conceptual framework for understanding evolving human-AI relationships, but its real value emerges through implementation. Effective implementation begins with *continuous dimensional positioning* and monitoring throughout a system's lifecycle. Rather than one-time classification, organisations must track how systems evolve. By monitoring user reliance patterns, system self-modifications, and responsibility distributions, organisations can detect when systems approach critical thresholds before risks materialise.

In practice, organisations could implement dimensional tracking through several concrete approaches. A healthcare institution deploying diagnostic AI might establish regular audits that measure clinician override rates (tracking de facto authority shifts), monitor system response times to unexpected inputs (measuring autonomous adaptation), and document responsibility handoffs between technical and medical teams (mapping accountability distribution). Similarly, financial institutions could create dimensional dashboards that visualise how trading algorithms' decision boundaries evolve over time, highlighting when systems approach predefined thresholds that trigger enhanced review protocols. These lightweight tracking mechanisms allow organisations to detect subtle governance shifts before they create regulatory or safety concerns, without requiring wholesale replacement of existing oversight structures.

*Thresholds serve as governance pivots*, triggering calibrated oversight responses. When systems cross the *Verification-to-Delegation* threshold, governance should shift toward statistical auditing and enhanced exception handling. Similarly, crossing the *Process-to-Outcome* threshold requires pivoting from transparency mechanisms toward outcome validation. Systems approaching the *Information-to-Authority* threshold require stronger override capabilities and formalised accountability protocols.

Sustainable governance ultimately depends on *dynamic trust calibration*. Different applications warrant different trust profiles—medical systems demand higher accountability standards than entertainment recommendations. Organisations should develop trust maps identifying where stakeholder expectations align with technical performance to maintain legitimacy.

Importantly, dimensional governance *complements rather than replaces* categorical frameworks. A hybrid approach leverages categories as regulatory anchors while using dimensions for continuous calibration. For instance, risk-tier classifications in the EU AI Act could establish baseline requirements, while dimensional tracking enables proportional oversight within those boundaries. This balances regulatory clarity with operational adaptability in a cohesive framework that can evolve alongside advancing AI capabilities.

# 5. Conclusion: Governing for a Dynamic Future

The era of agentic AI demands rethinking governance frameworks. Existing categorical models risk becoming outdated not because categories themselves are problematic, but because static definitions fail to adapt to evolving capabilities. What is needed is an adaptive definition of categories based on underlying dimensions like *the 3As* we propose. By tracking AI systems' movement along these dimensions, we create oversight architectures that are responsive, anticipatory, and stakeholder-sensitive.

Dimensional governance recognises that different communities—legal scholars, technologists, ethicists, and policymakers—may prioritise different facets of AI evolution. This approach allows each to engage with aspects most relevant to their expertise while coordinating within a shared framework, making governance a collaborative ecosystem rather than a mere classification exercise.

Critics may question whether such flexibility is practical. The concern about regulatory certainty is crucial but rests on a particular conception of what certainty means. Traditional categorical regulation provides certainty through fixedness; dimensional governance provides certainty through transparency of process. Rather than knowing "my system is permanently low-risk," organisations would know "these are the measured dimensions, these are the current thresholds, and this is the evidence-based process through which they may evolve." We acknowledge implementation challenges, particularly in quantifying systems' positions along each dimension and developing standardised measurement methodologies, but these are challenges of refinement rather than fundamental barriers. The question is not whether to have certainty, but which form of certainty better serves rapidly evolving technologies.

By explicitly acknowledging the continuous dimensions underlying AI systems' behaviour, we provide a more robust foundation upon which necessary regulatory categories can be built. Risk tiers and oversight models remain essential for practical governance but become more effective when understood as deliberate thresholds across underlying dimensions.

Ultimately, dimensional governance prepares us not just for what AI systems are today, but for what they are becoming—designing governance that evolves alongside the very systems it stewards. In a future shaped by increasingly agentic AI, such adaptability will not be optional; it will be essential.

**Acknowledgements:** We are grateful to the anonymous reviewers for their thoughtful and constructive feedback, which substantially improved this Commentary. Their insightful questions and suggestions helped us clarify key concepts and strengthen our arguments.

**Data availability statement:** This theoretical framework paper does not involve empirical data collection or analysis. All sources are publicly available academic publications and policy documents as cited in the references. No datasets or supplementary materials were created for this study.

**Declaration of interests:** ZE is an Editor-in-Chief of the Data & Policy journal and Founding Director of Data for Policy CIC, which provides editorial support to Data & Policy. DH is also a Director at Data for Policy CIC.

**Funding:** This research did not receive any specific grant from funding agencies in the public, commercial, or not-for-profit sectors.

**Declaration of AI use:** AI tools were used to refine language and presentation for clarity. The authors take full responsibility for the content of the article.